# A Public Website for the Automated Assessment and Validation of SARS-CoV-2 Diagnostic PCR Assays

Po-E Li[1], Adán Myers y Gutiérrez[1], Karen Davenport, Mark Flynn, Bin Hu, Chien-Chi Lo, Elais Player Jackson, Migun Shakya, Yan Xu, Jason Gans*, and Patrick S. G. Chain*

Bioscience Division, Los Alamos National Laboratory, Los Alamos, New Mexico

[1]Contributed equally to this work.

*To whom correspondence should be addressed.

## Abstract

**Summary:** Polymerase chain reaction-based assays are the current gold standard for detecting and diagnosing SARS-CoV-2. However, as SARS-CoV-2 mutates, we need to constantly assess whether existing PCR-based assays will continue to detect all known viral strains. To enable the continuous monitoring of SARS-CoV-2 assays, we have developed a web-based assay validation algorithm that checks existing PCR-based assays against the ever-expanding genome databases for SARS-CoV-2 using both thermodynamic and edit-distance metrics. The assay screening results are displayed as a heatmap, showing the number of mismatches between each detection and each SARS-CoV-2 genome sequence. Using a mismatch threshold to define detection failure, assay performance is summarized with the true positive rate (recall) to simplify assay comparisons.
**Availability:** https://covid19.edgebioinformatics.org/#/assayValidation
**Contact:** Jason Gans (jgans@lanl.gov) and Patrick Chain (pchain@lanl.gov)

## 1 Introduction

Many aspects of the control, management and treatment responses to the global COVID-19 pandemic require accurate detection of its causative agent, SARS-CoV-2. To address this challenge, research groups around the world have developed Polymerase Chain Reaction (PCR)-based assays to detect SARS-CoV-2 genomic RNA (Supplementary Tables S1).

The impact of SARS-CoV-2 genetic drift on the ability of PCR-based assays to successfully detect target sequences is a concern. To address this concern, we have developed a web-based application that monitors existing SARS-CoV-2 PCR-based assays that are in use around the world and provides a visual summary of assay performance. Both the acquisition of new genomes and the assay validation process is automated, so that assays are checked and displayed daily to give near real-time results.

## 2 Implementation

The core of the validation algorithm is the ThermonucleotideBLAST (Gans and Wolinsky, 2008) *in silico* PCR screening tool. Publicly available assays are used as queries in ThermonucleotideBLAST and searched against a target database of SARS-CoV-2 genomes from the Global Initiative on Sharing All Influenza Data (GISAID) (Shu and McCauley, 2017) and Genbank (Clark et al., 2016).

Sequences are downloaded daily from these databases and filtered to exclude any that are less than 29 kilobases or are pangolin-SARS and bat-SARS. For sequences found in both databases, only the GenBank version is retained. True positives are defined as any assay-target pairwise alignment that contains in 0, 1, or 2 mismatches (in the oligonucleotide with the most mismatches). False negatives are defined as an assay/target combination that has either (a) one or more oligo/target pairwise alignments with 3 or more mismatches, or (b) one or more predicted oligo/target melting temperatures less than 40°C. Since all of the included assays are intended to detect SARS-CoV-2 and false positives are not predicted, assay performance is quantified by the recall (defined as the number of true positives divided by the sum of true positives and false negatives).

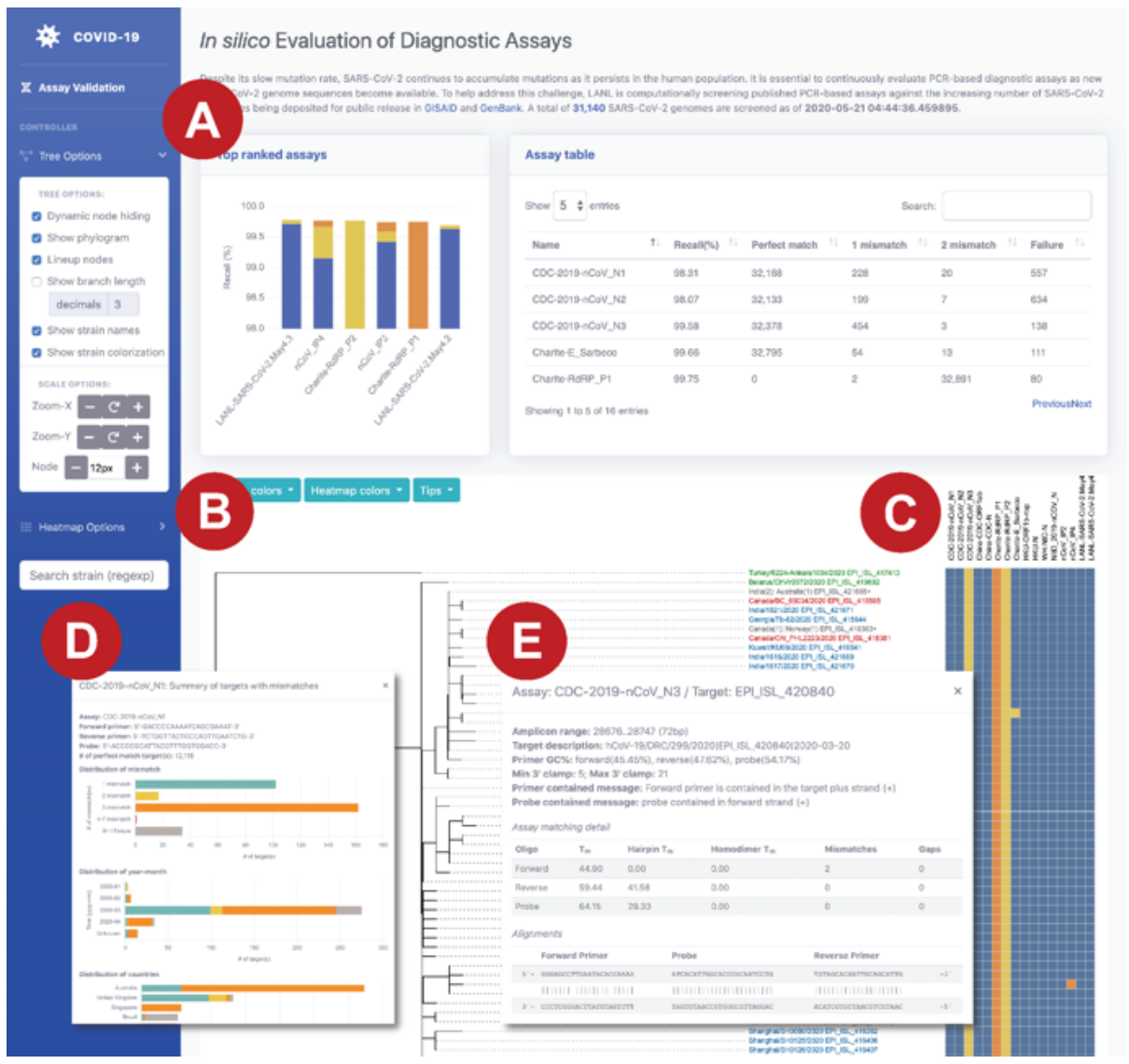

**Figure 1. Visualization of *in silico* evaluation of diagnostic assays.** (A) Dashboard including a bar chart and table with per-assay recall and mismatch counts; (B) Phylogenetic tree created from high-quality genomes color-labeled by continent; (C) A heatmap display of assay assessment per assay per genome; (D) Assay details and statistics of genomes with mismatches; (E) Detailed assay evaluation results, including alignments and thermodynamic information.

Per-assay recall values are summarized in the dashboard (Fig. 1A). The assays with the best recall rates are shown in a bar chart, which also displays detailed mismatch counts. The total mismatch and failure results are summarized in the per-assay table of aggregated data. Selecting any bar in the chart or assay in the table will display additional information on the distribution of targets with mismatches (Fig. 1D).

A phylogenetic tree (Fig. 1B), created using PhaME (Shakya et al. 2020), is derived from genomes defined by GISAID as high-quality (<1% Ns and <0.05% unique mutations). The leaves on the tree are represented by the genome labels and color-coded by geographic location. Mousing over the genome labels displays metadata associated with the sample. Identical SARS-CoV-2 sequences are clustered and represented as collapsed branches in the tree. The heatmap (Fig. 1C), color-coded to indicate the number of mismatches, shows analysis of every combination of assay and SARS-CoV-2 genome sequence. Selecting an individual cell of the heatmap displays detailed pairwise alignment information (Fig. 1E). This visualization is rendered using a custom PhyD3 phylogenetic tree viewer (Kreft et al., 2017).

## 3 Discussion

Few other public resources exist for assessing the performance of PCR-based SARS-CoV-2 assays. GISAID, one of the primary repositories for SARS-CoV-2 genomes, provides a high-level summary of PCR-based assay performance for registered users. However, this information is provided in the form of a static image with only a limited amount of information. The virological.org website provides static tables summarizing the high-level performance of PCR assays that have been periodically uploaded (Holland et al., 2020). Unlike these resources, the web-based application presented here provides a more detailed and interactive view of molecular assay performance that is updated regularly with recently deposited genomes (>31K as of May 22, 2020).

The heatmap-phylogeny view reveals patterns in predicted assay performance, including mismatches for the Charité RdRP assays (Vogels et al., 2020, Corman, *et al.,* 2020) that were originally developed for testing SARS and/or SARS-related bat coronaviruses (Fig. 1C). A different pattern, previously noted by Vogels et al. (Vogels et al., 2020), is seen within a subset of phylogenetically related strains due to a mismatch in the USA CDC N3 assay (CDC, 2020). Finally, a set of regularly updated assay designs are also included. Each of these assays is selected to individually maximize the predicted recall while maintaining multiplex compatibility with the other assays in the set. As genomics continues to be used for understanding pathogen outbreaks, resources such as the one provided by this website may help in the early identification of potential assay concerns, and provide guidance on alternate assay designs early on, to mitigate current assays that may be eroding.

## Acknowledgments

The authors declare no conflict of interest. Hosting of edgebioinformatics.org is provided by Cyverse, which is supported by the National Science Foundation under Award Numbers DBI-0735191, DBI-1265383, and DBI-1743442. We acknowledge the authors and originators of sequences from the submitting laboratories who have contributed to the GISAID database.

## Funding

This research was supported by LANL (20200732ER), by DTRA (CB10152 and CB10623) and by the DOE Office of Science (KP160101), through the National Virtual Biotechnology Laboratory, a consortium of DOE national laboratories focused on response to COVID-19, with funding provided by the Coronavirus CARES Act.

**Supplementary Table S1**: Publicly available assays and the sequences of the assay oligonucleotides

| Name | 5-Forward-3’ | 5’-Reverse-3’ | 5’-Probe-3’ | Source (See https://www.who.int/emergencies/diseases/novel-coronavirus-2019/technical-guidance/laboratory-guidance) |
|---|---|---|---|---|
| CDC-2019-nCoV_N1 | GACCCCAAAATCAGCGAAAT | TCTGGTTACTGCCAGTTGAATCTG | ACCCCGCATTACGTTTGGTGGACC | https://www.cdc.gov/coronavirus/2019-ncov/lab/rt-pcr-panel-primer-probes.html |
| CDC-2019-nCoV_N2 | TTACAAACATTGGCCGCAAA | GCGCGACATTCCGAAGAA | ACAATTTGCCCCCAGCGCTTC | https://www.cdc.gov/coronavirus/2019-ncov/lab/rt-pcr-panel-primer-probes.html |
| CDC-2019-nCoV_N3 | GGGAGCCTTGAATACACCAAAA | TGTAGCACGATTGCAGCATTG | A**Y**CACATTGGCACCCGCAATCCTG | https://www.cdc.gov/coronavirus/2019-ncov/lab/rt-pcr-panel-primer-probes.html |
| China-CDC-ORF1ab | CCCTGTGGGTTTTACACTTAA | ACGATTGTGCATCAGCTGA | CCGTCTGCGGTATGTGGAAAGGTTATGG | https://www.nejm.org/doi/10.1056/NEJMoa2001017 |
| China-CDC-N | GGGGAACTTCTCCTGCTAGAAT | CAGACATTTTGCTCTCAAGCTG | TTGCTGCTGCTTGACAGATT | https://www.nejm.org/doi/10.1056/NEJMoa2001017 |
| Charite-RdRP_P1 | GTGA**R**ATGGTCATGTGTGGCGG | CA**R**ATGTTAAA**S**ACACTATTAGCATA | CCAGGTGG**W**AC**R**TCATC**M**GGTGATGC | https://www.who.int/docs/default-source/coronaviruse/protocol-v2-1.pdf?sfvrsn=a9ef618c_2 |
| Charite-RdRP_P2 | GTGA**R**ATGGTCATGTGTGGCGG | CA**R**ATGTTAAA**S**ACACTATTAGCATA | CAGGTGGAACCTCATCAGGAGATGC | https://www.who.int/docs/default-source/coronaviruse/protocol-v2-1.pdf?sfvrsn=a9ef618c_2 |
| Charite-E_Sarbeco | ACAGGTACGTTAATAGTTAATAGCGT | ATATTGCAGCAGTACGCACACA | ACACTAGCCATCCTTACTGCGCTTCG | https://www.who.int/docs/default-source/coronaviruse/protocol-v2-1.pdf?sfvrsn=a9ef618c_2 |
| HKU-ORF1b-nsp | TGGGG**Y**TTTAC**R**GGTAACCT | AAC**R**CGCTTAACAAAGCACTC | TAGTTGTGATGC**W**ATCATGACTAG | https://www.who.int/docs/default-source/coronaviruse/peiris-protocol-16-1-20.pdf?sfvrsn=af1aac73_4 |
| HKU-N | TAATCAGACAAGGAACTGATTA | CGAAGGTGTGACTTCCATG | GCAAATTGTGCAATTTGCGG | https://www.who.int/docs/default-source/coronaviruse/peiris-protocol-16-1-20.pdf?sfvrsn=af1aac73_4 |
| WH-NIC-N | CGTTTGGTGGACCCTCAGAT | CCCCACTGCGTTCTCCATT | CAACTGGCAGTAACCA | https://www.who.int/docs/default-source/coronaviruse/conventional-rt-pcr-followed-by-sequencing-for-detection-of-ncov-rirl-nat-inst-health-t.pdf?sfvrsn=42271c6d_4 |
| NIID_2019-nCOV_N | AAATTTTGGGGACCAGGAAC | TGGCACCTGTGTAGGTCAAC | ATGTCGCGCATTGGCATGGA | https://www.jstage.jst.go.jp/article/yoken/advpub/0/advpub_JJID.2020.061/_pdf |
| nCoV_IP2 | ATGAGCTTAGTCCTGTTG | CTCCCTTTGTTGTGTTGT | AGATGTCTTGTGCTGCCGGTA | https://www.who.int/docs/default-source/coronaviruse/real-time-rt-pcr-assays-for-the-detection-of-sars-cov-2-institut-pasteur-paris.pdf?sfvrsn=3662fcb6_2 |
| nCoV_IP4 | GGTAACTGGTATGATTTCG | CTGGTCAAGGTTAATATAGG | TCATACAAACCACGCCAGG | https://www.who.int/docs/default-source/coronaviruse/real-time-rt-pcr-assays-for-the-detection-of-sars-cov-2-institut-pasteur-paris.pdf?sfvrsn=3662fcb6_2 |

## Discussion Addendum

While there are many sources of empirical assay validation, only a few current resources exist that present predictive assessments. Online repositories and message groups now offer the means for presenting and sharing information related to the performance of current assays. This information can take the form of static analyses (Which presumably employ *in silico* predictive models for performance assessment) or simply sharing of information (often in manuscript format) in a workspace or topic group chat.

GISAID (Shu, *et al.,* 2017), in addition to being a repository for SAR-CoV-2 genomes, supplies bioinformatic analyses of the database and a summary of assay performance. The assay assessment format is a static slide with summary information. The nature of the content supplied here has varied over time. Early, the slide showed selected assays and listed genomes with mismatches to the assay (also indicating the specific mismatches). As the number of genomes has grown, the list of genomes on this slide grew, while losing the specific mismatch information. This was subsequently replaced with a list of assays and their mismatches with respect to a reference genome. The current display shows a histogram of the percent of genomes with a mutation in the primer region for each assay and another histogram with the percent of genomes with a mutation in the five 3' terminal nucleotides. This analysis is developed from BLASTN searches. In general, GISAID offers useful analysis, but the content has varied over time and the presentation is static. Updates to the analysis are frequent but unpredictable.

In contrast, virological.org provides a forum for discussion of virus bioinformatics analysis. Information may be presented in a static format, such as a manuscript. While this enables conversation on the topic, the analysis is static, and information is only updated as often as posters choose to upload to the site.

Our web-based application provides an alternative that is interactive and updated daily with recently deposited genomes. Our analysis uses an *in silico* algorithm that is predictive of assay performance. A glance at the heatmap will quickly reveal patterns and trends in the performance of the assays.

It is immediately evident that two of the Charité assays for RdRP are imperfect matches to all strains of SARS-CoV-2 (Fig. S1). Figures S2 and S3 show the alignments of the probe 1 and probe 2 assay (respectively) to Turkey/6224-Ankara1034 genome as an example. It is evident that both assays have one consistent mismatch in the reverse primer and the P1 assay has two mismatches in the probe. These mismatches occur on all genomes visualized. The alignment shows that the Charité reverse primer has a G or a C opposite a T in the genome. Evidently, the Charité assay has been designed with a degenerate base at this position where the SARS-CoV-2 genome appears to consistently have a T. This design appears to be a relic of the fact that this primer was initially designed for testing for SARS or bat-SARS coronaviruses (Vogels et al., 2020).

Another pattern that is evident from the heatmap (Fig. S1) is that the N3 assay from USA CDC has a number of strains with a single mismatch as well. This is a T base in the forward primer that mismatches with a G on the target genome (Vogels et al., 2020) (Fig S4). While most strains of virus do not have this mismatch with the assay at this base, a certain number of strains appear to have acquired this mutation. The tree reveals that there is a clade with 25 strains that have acquired this mutation resulting in the mismatch to the N3 assay (Fig. S5). Most of these strains originate from Asia and Oceania, with a few in North American and Europe. The majority of these strains were collected in early March, corroborating a common origin for this clade. The remainder of strains with this mutation do not appear to share an obviously delineated clade (Fig. S6), but many of these originate from Asia and Oceania and many have a collection date of late February and early March (with some as early as January) suggesting an earlier origin for the mutation in this group.

Other small clusters of mutations abound. There is a small clade of 23 sequences that cause failure of the N1 assay due to 3 mismatches on the probe (Fig. S7). These bases (ACC) are on the first three bases of the probe opposite TGA (Fig. S8). While C-A is a literal mismatch, the remaining two bases should not pair due to the thermodynamic unfavorability of their position on the end of an oligo. The majority of these sequences originate in Oceania in early March. There does not appear to be nearby clades with mismatches, so this group appears to have emerged fairly rapidly. This is the type of development that will potentially lead to obsolescence of assays, and so the ability to track these types of events is particularly valuable.

Here we have presented a web-based tool for visual and quantitative predictive assessment and validation of PCR assays that are designed to detect SARS-CoV-2. The tool we have described provides an easy-to-read visual representation of assay performance with respect to the target genomes and the evolutionary relationship between the genomes. This web application will thus offer utility to policy makers, health officials, and other stakeholders in making decisions about COVID-19 testing and tracking of the SARS-CoV-2 virus.

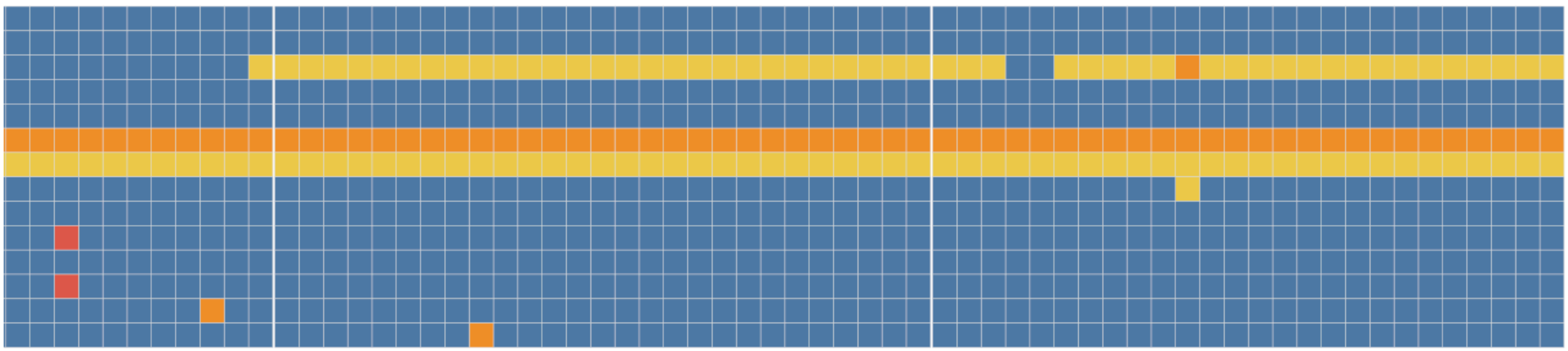

Figure S1.

| | Forward Primer | Probe | Reverse Primer | |
|---|---|---|---|---|
| 5'– | GTGAAATGGTCATGTGTGGCGG | CCAGGTGGAACATCATCAGGTGATGC | CAAATGTTAAAGACACTATTAGCATA | –3' |
| | \|\|\|\|\|\|\|\|\|\|\|\|\|\|\|\|\|\|\|\|\|\| | \|\|\|\|\|\|\|\|\|\|\| \|\|\|\|\|\|\|\| \|\|\|\|\| | \|\|\|\|\|\|\|\|\|\|\| \|\|\|\|\|\|\|\|\|\|\|\|\|\| | |
| 3'– | CACTTTACCAGTACACACCGCC | GGTCCACCTTGGAGTAGTCCTCTACG | GTTTACAATTTTGTGATAATCGTAT | –5' |

Figure S2

| | Forward Primer | Probe | Reverse Primer | |
|---|---|---|---|---|
| 5'– | GTGAAATGGTCATGTGTGGCGG | CAGGTGGAACCTCATCAGGAGATGC | CAAATGTTAAAGACACTATTAGCATA | –3' |
| | \|\|\|\|\|\|\|\|\|\|\|\|\|\|\|\|\|\|\|\|\|\| | \|\|\|\|\|\|\|\|\|\|\|\|\|\|\|\|\|\|\|\|\|\|\|\|\| | \|\|\|\|\|\|\|\|\|\|\| \|\|\|\|\|\|\|\|\|\|\|\|\|\| | |
| 3'– | CACTTTACCAGTACACACCGCC | GTCCACCTTGGAGTAGTCCTCTACG | GTTTACAATTTTGTGATAATCGTAT | –5' |

Figure S3

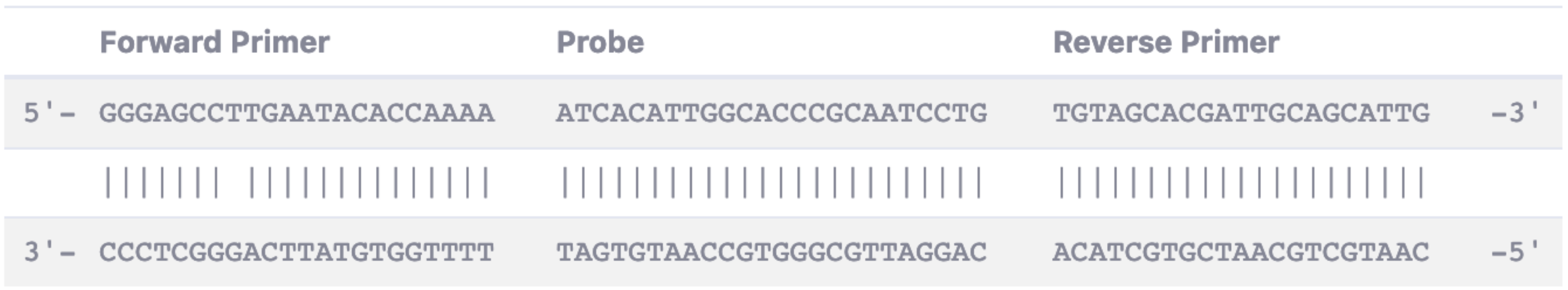

Figure S4

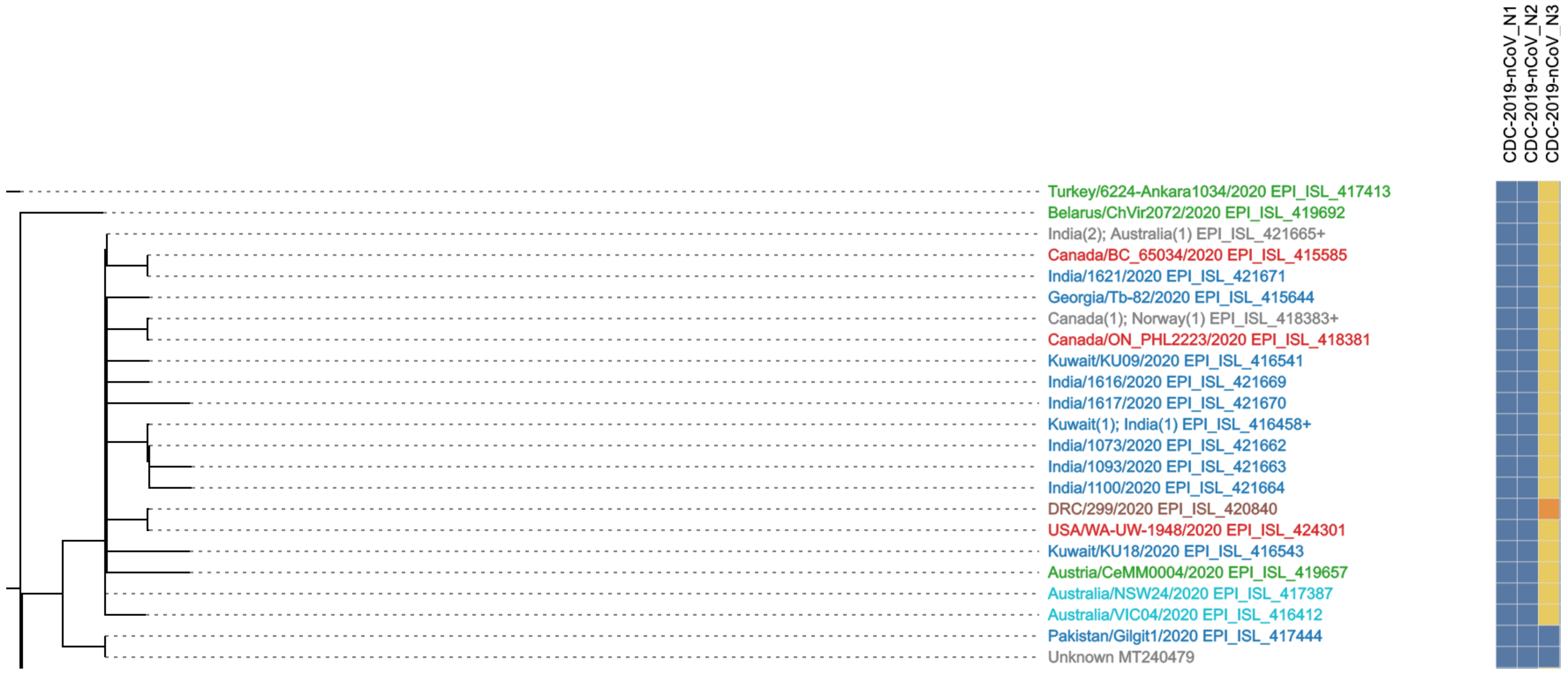

Figure S5

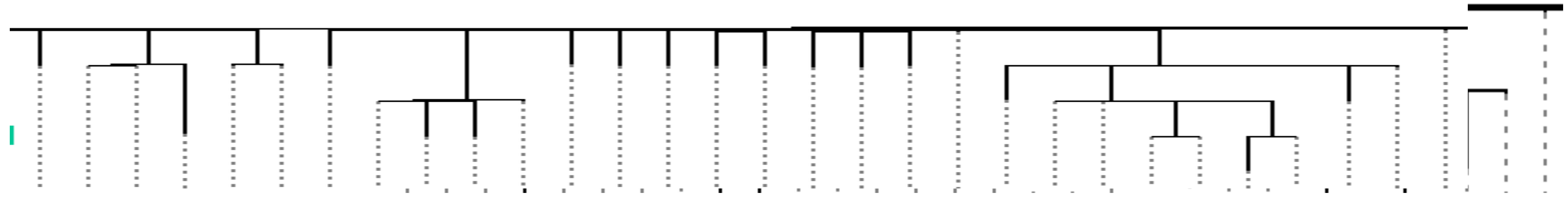

Figure S6

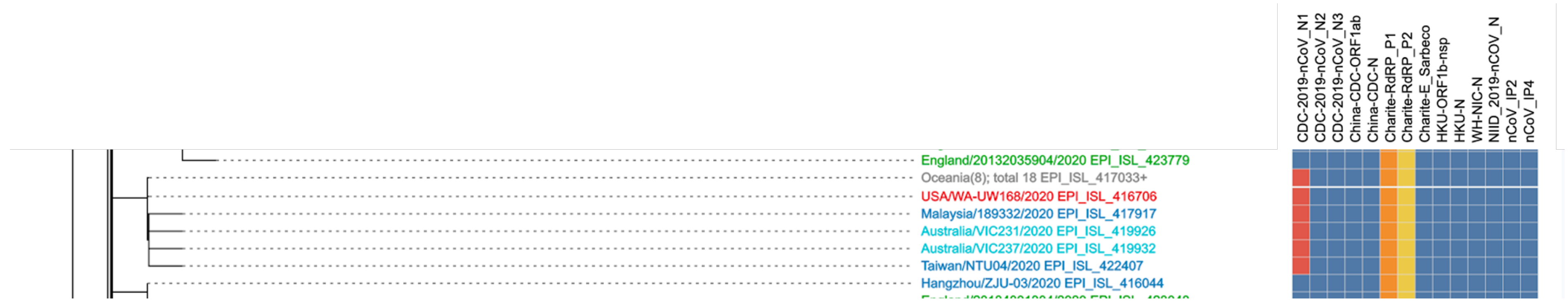

Figure S7

| | Forward Primer | Probe | Reverse Primer | |
|---|---|---|---|---|
| 5'– | GACCCCAAAATCAGCGAAAT | ACCCCGCATTACGTTTGGTGGACC | TCTGGTTACTGCCAGTTGAATCTG | –3' |
| | |||||||||||||||||||| | ||||||||||||||||||||| | |||||||||||||||||||||||| | |
| 3'– | CTGGGGTTTTAGTCGCTTTA | TGAGGCGTAATGCAAACCACCTGG | AGACCAATGACGGTCAACTTAGAC | –5' |

Figure S8